\documentclass[reprint,twocolumn,aps,prl,amsmath,amssymb,floatfix,superscriptaddress]{revtex4-1}

\usepackage{graphicx}
\usepackage{epsfig}
\usepackage{bm}
\usepackage{dcolumn}
\usepackage{color}
\usepackage[colorlinks,urlcolor=cyan,citecolor=blue,linkcolor=magenta]{hyperref}
\makeatletter
\newcommand*{\rom}[1]{\expandafter\@slowromancap\romannumeral #1@}
\makeatother

\begin{document}

\title{Angular Stripe Phase in Spin-Orbital-Angular-Momentum Coupled Bose Condensates}

\author{Xiao-Long Chen}
\affiliation{Centre for Quantum and Optical Science, Swinburne University of Technology,
Melbourne, Victoria 3122, Australia}

\author{Shi-Guo Peng}
\email{pengshiguo@wipm.ac.cn}
\affiliation{State Key Laboratory of Magnetic Resonance and Atomic and Molecular Physics, Wuhan Institute of Physics and Mathematics, Chinese Academy of Sciences, Wuhan, 430071, China}

\author{Peng Zou}
\email{phy.zoupeng@gmail.com}
\affiliation{College of Physics, Qingdao University, Qingdao 266071, China}

\author{Xia-Ji Liu}
\affiliation{Centre for Quantum and Optical Science, Swinburne University of Technology,
Melbourne, Victoria 3122, Australia}

\author{Hui Hu}
\affiliation{Centre for Quantum and Optical Science, Swinburne University of Technology,
Melbourne, Victoria 3122, Australia}

\date{\today}
\begin{abstract}
We propose that novel superfluid with supersolid-like properties - angular stripe phase - can be realized in a pancake-like spin-1/2 Bose gas with spin-orbital-angular-momentum coupling. We predict a rich ground-state phase diagram, including the vortex-antivortex pair phase, half-skyrmion phase, and two different angular stripe phases. The stripe phases feature modulated angular density-density correlation with sizable contrast and can occupy a relatively large parameter space. The low-lying collective excitations, such as the dipole and breathing modes, show distinct behaviors in different phases. The existence of the novel stripe phase is also clearly indicated in the energetic and dynamic instabilities of collective modes near phase transitions. Our predictions of the angular stripe phase could be readily examined in current cold-atom experiments with $^{87}$Rb and $^{41}$K.
\end{abstract}

\maketitle
{\color{blue}\emph{Introduction}}.---\textemdash Owing to the high controllability of degrees of freedom, ultracold atomic gases have became a versatile platform to study artificial gauge fields over the last few years~\cite{dalibard2011colloquium,goldman2014light}. A prominent example is the spin-orbit coupling (SOC), the coupling between a particle's spin and momentum, which plays a crucial role in many fascinating phenomena such as the quantum spin Hall effect and topological superfluidity~\cite{Galitski2013,zhai2015degenerate}. Theoretically, an exotic phase of matter, namely the stripe phase, was predicted to exist in a Bose condensate with Rashba-type SOC~\cite{wang2010spin,cong2011unconventional} or Raman-laser-induced SOC~\cite{ho2011bose,li2012quantum}. In analogy to the long-sought supersolid phase in solid helium~\cite{boninsegni2012colloquium}, it breaks both continuous translational symmetry to form a crystalline pattern and $U(1)$ gauge symmetry with atoms moving frictionlessly as in a superfluid. The stripe phase is energetically favored for the intra-spin interaction strength larger than the inter-spin one (e.g., $g>g_{_{\uparrow\downarrow}}$), and in a $^{87}$Rb gas with Raman SOC, it lies in a narrow window of Rabi frequency $\Omega$ due to the small difference between $g$ and $g_{_{\uparrow\downarrow}}$~\cite{li2012quantum}. This is to be compared with the plane-wave and zero-momentum phases appearing at larger Rabi frequency. As a result, a direct observation of the stripe phase remains elusive, due to the short period and negligible density contrast of the stripes~\cite{lin2011spin,ji2014experimental,martone2014approach,chen2018quantum}. It has been indirectly probed using Bragg spectroscopy in a SOC Bose gas on lattices, where the interaction difference is effectively enhanced~\cite{li2017stripe}. 
\begin{figure}[t]
\centering
\includegraphics[width=0.48\textwidth]{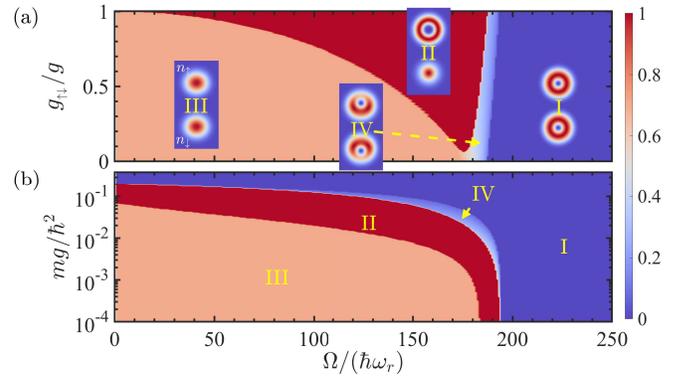}
\caption{Phase diagram of a harmonically trapped pancake-like SOAMC Bose gas with particle number $N=10^{3}$, in the $g_{_{\uparrow\downarrow}}/g$-$\Omega$ plane at $mg/\hbar^{2}=0.01$ (a) and in the $g$-$\Omega$ plane at $g_{_{\uparrow\downarrow}}/g=0.5$ (b). The color indicates the portion of the $l_{z}=\hbar$ state in the condensate wavefunction. The insets show typical densities for each spin states in different phases. The phase boundaries are discussed in the text.}
\label{fig1_phase}
\end{figure}

In addition to the Raman SOC, a new type of SOC, the so-called spin-orbital-angular-momentum coupling (SOAMC), has been recently proposed in pioneering theoretical works~\cite{hu2015half,demarco2015angular,qu2015quantum,sun2015spin,chen2016spin}: by utilizing two co-propagating Laguerre-Gaussian (LG) laser beams to induce an \emph{off-diagonal} Stark shift, the atomic pseudospin can be coupled to its angular momentum. The resulting SOAMC is of a two-dimensional (2D) nature with an axial symmetry, which gives rise to intriguing quantum phases, such as the vortex-antivortex pair phase (see the phase \rom{1} in Fig.~\ref{fig1_phase}) with definite angular momentum $l_{z}=0$, the half-skyrmion phase \rom{2} with $l_{z}=\hbar$ or $-\hbar$, and the superposition phase \rom{3} with an equal-weight combination of two angular-momentum states at $l_{z}=\pm\hbar$~\cite{hu2015half,demarco2015angular}. These are precisely the analogues of the zero-momentum, plane-wave and stripe phases in the case of Raman SOC. For this reason, the phase \rom{3} has been previously referred to as the stripe phase~\cite{demarco2015angular,qu2015quantum}. The SOAMC was most recently engineered in spin-1~\cite{chen2018spin,chen2018rotating} and spin-$1/2$ $^{87}$Rb Bose gases~\cite{zhang2019ground}. Different from the theoretical proposals~\cite{hu2015half,demarco2015angular,qu2015quantum,sun2015spin,chen2016spin}, LG laser beams with tune-out wavelength $\lambda=790.02$nm were used to eliminate the unnecessary \emph{diagonal} Stark shift~\cite{chen2018spin,chen2018rotating,zhang2019ground}. In this work, we show that this slight experimental improvement may lead to a relatively broad stripe phase and another stripe phase which is less sensitive to the interaction difference $g-g_{_{\uparrow\downarrow}}$, see for example, the phases \rom{3} and \rom{4} in Fig.~\ref{fig1_phase}, which will be collectively named as \emph{angular} stripe phase.

\begin{figure}[t]
\centering
\includegraphics[width=0.48\textwidth]{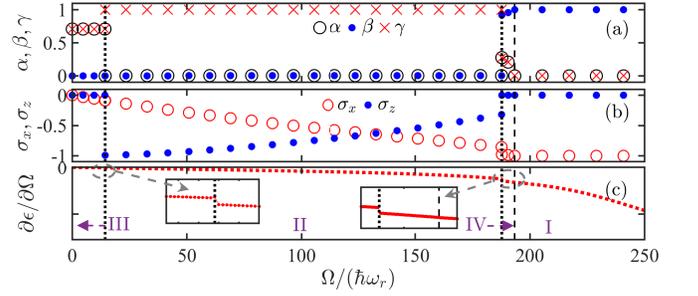}
\caption{The weighting coefficients $\alpha$, $\beta$, $\gamma$ (a), spin polarizations $\langle\sigma_{x}\rangle$ and $\langle\sigma_{z}\rangle$ (b), and the derivative of mean-field energy $\partial\epsilon/\partial\Omega$, as a function of $\Omega$ at $g=0.01\hbar^{2}/m$, for a pancake SOAMC $^{87}$Rb Bose gas with $g_{_{\uparrow\downarrow}}/g=100.40/100.86$ and $N=10^{3}$. The vertical black dotted (dashed) lines indicate the first-order (second-order) phase transitions.}
\label{fig2_coeff}
\end{figure}

\begin{figure}[t]
\centering{}
\includegraphics[width=0.48\textwidth]{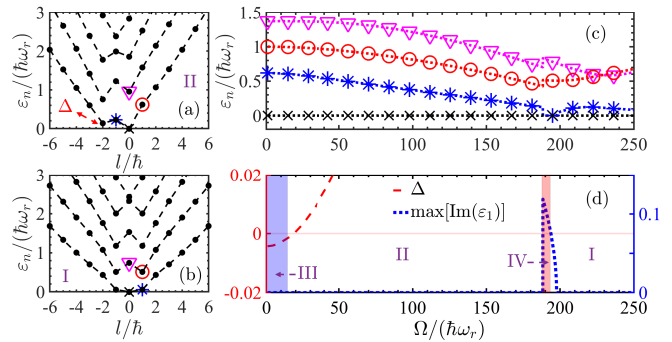}
\caption{The excitation spectrum as a function of the angular momentum $l$ of Bogoliubov quasiparticles at $\Omega/(\hbar\omega_{r})=150$ (a) and $200$ (b). The roton minimum is denoted by $\Delta$. In (c) and (d), the mode frequencies $\varepsilon_{n}(l)$, the roton energy $\Delta$ and the maximum imaginary part of the lowest branch $\varepsilon_{n=1}(l)$, are shown as a function of Rabi frequency $\Omega$. The blue and red shaded areas highlight the regimes of the angular stripe phases (\rom{3}) and (\rom{4}), respectively, as determined from Fig.~\ref{fig2_coeff}. The other parameters are the same as in Fig.~\ref{fig2_coeff}.}
\label{fig3_collective}
\end{figure}

\begin{figure}[t]
\centering{}
\includegraphics[width=0.48\textwidth]{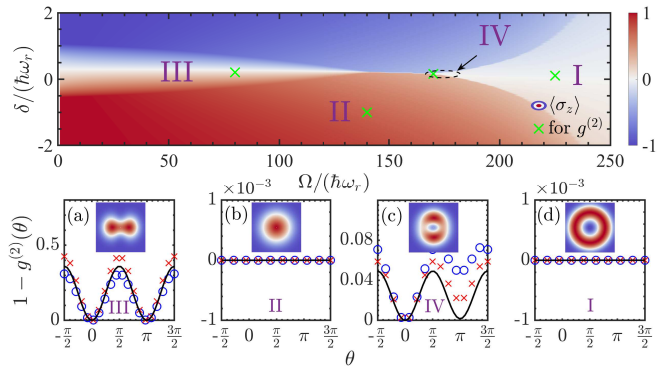}
\caption{Phase diagram in the $\delta$-$\Omega$ plane of a SOAMC $^{41}$K gas, in harmonic traps $(\omega_r,\omega_z)=2\pi\times(10,200)$Hz with $N=10^{3}$ and $(a_{_{\uparrow\uparrow}},a_{_{\downarrow\downarrow}},a_{_{\uparrow\downarrow}})=(65,100,10)a_0$. The color indicates the expectation value of spin magnetization $\langle\sigma_z\rangle$ in the condensate wavefunction. In (a-d), the angular density-density correlation function $g^{(2)}(\theta)$ is shown for different phases (green crosses in top figure). The results for spin-up, spin-down and total density are indicated by circles, crosses and solid lines, respectively. The insets show the total density profiles.} 
\label{fig4_n_correlation}
\end{figure}

Quite generally, the angular stripe phase favors a \emph{double} or \emph{multiple} occupation of the angular-momentum states that are energetically allowed (see Fig.~\ref{fig2_coeff}(a) and Eq.~\eqref{eq:2D-variationalansatz}) and explicitly breaks the axial symmetry of the condensate wavefunction. It thus possesses spatial modulation in the angular direction. This is best manifested in the angular density-density correlation function $g^{(2)}(\theta)$ for each spin state, which clearly reveals distinguishable symmetries in spatial density distributions (Fig.~\ref{fig4_n_correlation}(a-d)). We find a relatively large oscillation period and a sizable contrast in the oscillation amplitude, making a direct experimental detection of the angular stripe feasible. We also consider the low-lying collective excitations such as the dipole and breathing modes in different quantum phases and investigate the Rabi frequency dependence of the mode frequencies (Fig.~\ref{fig3_collective}). The violation of Galilean invariance of this SOAMC system is demonstrated in the dipole mode. Moreover, the appearance of the angular stripe phases is evident from the energetic and dynamical instabilities of the excitation spectrum.

{\color{blue}\emph{The model}}.\textemdash We start by describing a setup similar to that in recent rubidium experiment~\cite{zhang2019ground}, where a pair of LG beams with different orbital angular momentum ($n_{1}=-2$ and $n_{2}=0$) co-propagate along the $z$ axis and generate the SOAMC in the $x$-$y$ plane. Two hyperfine states in the $F=1$ ground-state manifold of $^{87}$Rb are selected to act as pseudospin states $\left|\uparrow\right\rangle $ and $\left|\downarrow\right\rangle $. For convenience, we consider a quasi-2D configuration with a highly oblate harmonic trapping potential along the $z$-direction, which allows us to tune the overall effective strength of the inter-atomic interactions and to enlarge the phase space for the angular stripe phase. After a unitary transformation $\mathcal{U}=\mathrm{exp}(-in\phi\hat{\sigma}_{z})$ with $n\equiv(n_{1}-n_{2})/2$, in the polar coordinates $(r,\phi)$ the resulting pancake-like Bose gas with SOAMC at zero laser detuning $\delta=0$ can be described by a reduced model Hamiltonian $\mathcal{H}=\mathcal{H}_{s}+\mathcal{H}_{\mathrm{int}}$, where the single-particle part reads ~\cite{zhang2019ground}
\begin{equation} \label{eq:single-particle}
\mathcal{H}_{s}= -\frac{\hbar^2}{2mr}\partial_r\left(r\partial_r\right)+\frac{(\hat{L}_z-n\hbar\hat{\sigma}_z)^2}{2mr^2}+V_\mathrm{ext}+\Omega(r)\hat{\sigma}_x,
\end{equation}
$\hat{\sigma}_{x,z}$ are Pauli matrices, $V_{\mathrm{ext}}(r)=m\omega_{r}^{2}r^{2}/2$ is the trapping potential in the $x$-$y$ plane, and $\Omega(r)=\Omega(r/R)^{2}\mathrm{exp}[-2(r/R)^{2}]$ is the spatial-dependent coupling strength of laser beams with the Rabi frequency $\Omega$ and waist $R$. The canonical angular momentum operator $\hat{L}_{z}\equiv-i\hbar\partial_{\phi}$ is thus coupled to atomic spin via the SOAMC term $\sim\hat{L}_{z}\hat{\sigma}_{z}$. Hereafter, we use the characteristic energy $\hbar\omega_{r}$, oscillator length $d=\sqrt{\hbar/(m\omega_{r})}$ of the trap and $k_{r}\equiv1/d$ as the units of energy, length and wavevector, respectively. We also take a relatively large waist of laser beams $R=20d$ as in the experiments~\cite{chen2018spin,zhang2019ground}.

In the absence of inter-atomic interactions, atoms occupy into the single-particle states with a definite angular momentum $l_{z}$ such as $l_{z}=0$ or $\pm\hbar$, see Refs.~\cite{zhang2019ground,SM}. Taking into account interactions, at zero temperature the normalized spinor ground-state wavefunction ${\bf \Psi}\equiv\left(\Psi_{\uparrow},\Psi_{\downarrow}\right)$ can be determined by either minimizing the mean-field energy per particle $\epsilon=\iint r\mathrm{d}r\mathrm{d}\phi \Big[\left(\Psi_{\uparrow}^{*},\Psi_{\downarrow}^{*}\right)\mathcal{H}_{s}\left(\begin{array}{c}
\Psi_{\uparrow}\\
\Psi_{\downarrow}
\end{array}\right)+\frac{g_{_{\uparrow\uparrow}}}{2}|\Psi_{\uparrow}|^4+\frac{g_{_{\downarrow\downarrow}}}{2}|\Psi_{\downarrow}|^4+g_{_{\uparrow\downarrow}}|\Psi_{\uparrow}|^2|\Psi_{\downarrow}|^2\Big]/N$, or self-consistently solving the Gross-Pitaevskii equation (GPE)~\cite{SM,popov1991,griffin1996conserving,hu2012spin}. The latter approach is particularly useful when the condensate wavefunction $\psi_{_{l_{z}}}(r,\phi)=[\varphi_{_{l_{z}\uparrow}}(r),\varphi_{_{l_{z}\downarrow}}(r)]e^{il_{z}\phi}/\sqrt{2\pi}$ preserves the axial symmetry and the angular momentum $l_{z}$ remains as a good quantum number. In this case, we generalize the Bogoliubov theory to study low-energy elementary excitations~\cite{ramachandhran2012half,vasic2016excitation}, including dipole and breathing modes. The transitions between different phases may then be understood from possible instabilities of the collective modes.

When the axial symmetry is spontaneously breaking by interactions, we consider instead the energy-minimizing approach and adopt the following variational \textit{ansatz} for the condensate wavefunction~\cite{chen2016spin,SM}
\begin{equation}  \label{eq:2D-variationalansatz}
    {\bf \Psi}(r,\phi)=\alpha e^{i\theta_\alpha}\psi_{_{-1}}+\beta e^{i\theta_\beta}\psi_{_{0}}+\gamma e^{i\theta_\gamma}\psi_{_{1}},
\end{equation}
where $\psi_{_{l_{z}}}$ is the definite-angular-momentum state solved from the GPE with the interaction effect incorporated, $\alpha$, $\beta$, $\gamma$ are real non-negative weighting coefficients and $\theta_{\alpha}$, $\theta_{\beta}$, $\theta_{\gamma}$ are the phases. Thus, the mean-field energy $\epsilon$ becomes a functional of \emph{three} variational parameters, if we take into consideration the normalization condition $\alpha^{2}+\beta^{2}+\gamma^{2}=1$ and the fact that $\epsilon$ depends on the phase factors through a single function $\cos{(\theta_{\alpha}-2\theta_{\beta}+\theta_{\gamma})}$ only~\cite{SM}. The wavefunction ${\bf \Psi}$ is then determined from the minimization of $\epsilon$. We have checked that the use of more definite-angular-momentum states in the variational ansatz does not bring appreciable improvement for lowering $\epsilon$.

{\color{blue}\emph{Phase diagram}}.\textemdash In experiments with $^{87}$Rb and $^{23}$Na atoms in pancake traps~\cite{stock2005observation,clade2009observation,choi2013observation}, the interaction strengths $g_{_{\uparrow\uparrow}}=g_{_{\downarrow\downarrow}}=g\equiv\sqrt{8\pi}(a/a_{z})\hbar^{2}/m$ and $g_{_{\uparrow\downarrow}}\equiv\sqrt{8\pi}(a_{_{\uparrow\downarrow}}/a_{z})\hbar^{2}/m$ span approximately over the range $[0.01,0.15]\hbar^{2}/m$ and cross from the weakly-interacting to relatively strongly-interacting regimes. Here $a$ and $a_{_{\uparrow\downarrow}}$ are respectively the intra- and inter-spin $s$-wave scattering lengths in three dimensions and $a_{z}=\sqrt{\hbar/(m\omega_{z})}$ is the oscillator length along the tightly confined $z$-axis.

In Fig.~\ref{fig2_coeff}, we present the variational results for a Bose gas of $N=10^{3}$ $^{87}$Rb atoms with $g_{_{\uparrow\downarrow}}/g=a_{_{\uparrow\downarrow}}/a=100.40/100.86$~\cite{zhang2019ground} at a typical interaction strength $mg/\hbar^{2}=0.01$. By decreasing the Rabi frequency $\Omega$ from large values, we may identify four distinct regimes: (i) the vortex-antivortex pair phase \rom{1} with zero angular momentum $l_{z}=0$ (i.e., $\beta=1$, $\alpha=\gamma=0$), $\langle\sigma_{z}\rangle=0$ and $\langle\sigma_{x}\rangle=-1$; (ii) the angular stripe phase \rom{4} with no definite $l_{z}$ (i.e., $\beta\neq0$ and $\alpha=\gamma\neq0$), $\langle\sigma_{z}\rangle=0$ and $\langle\sigma_{x}\rangle\neq0$; (iii) the half-skyrmion phase \rom{2} with $l_{z}=-\hbar$ or $\hbar$ (i.e., $\alpha=1$, $\beta=\gamma=0$, or $\gamma=1$, $\alpha=\beta=0$), $\langle\sigma_{z}\rangle\neq0$ and $\langle\sigma_{x}\rangle\neq0$; and (iv) the angular stripe phase \rom{3}, which may be viewed as a specific case of the phase \rom{4} but with $\beta=0$ and $\alpha=\gamma=1/\sqrt{2}$. Compared with the results of Raman-induced or Rashba-type SOC~\cite{wang2010spin,li2012quantum}, it is readily seen that the phases \rom{3}, \rom{2} and \rom{1} are in \emph{one-to-one} correspondence with the well-known stripe, plane-wave and zero-momentum phases, respectively~\cite{demarco2015angular,qu2015quantum}. The novel angular stripe phase \rom{4} is unique to a SOAMC Bose gas, as a result of the discreteness of the angular momentum.

The nature of transition between different superfluid phases may be characterized by calculating the derivative of the mean-field energy with respect to the Rabi frequency~\cite{demarco2015angular,qu2015quantum}, $\partial\epsilon/\partial\Omega$, as shown in Fig.~\ref{fig2_coeff}(c) with vertical dashed and dotted lines for the second- and first-order transitions, respectively. We find a first-order transition from the angular stripe phase \rom{3} to the half-skyrmion phase \rom{2}, the same as the transition between the stripe and plane-wave phases in a Raman SOC Bose gas~\cite{li2012quantum,martone2012anisotropic}. The transition from the phase \rom{2} to the angular stripe phase \rom{4} is also of first order, accompanied by the sudden appearance of the component $\psi_{_{0}}$ in the condensate wavefunction ${\bf \Psi}$ (i.e., $\beta\neq0$) and sudden change in the spin polarizations. On the contrary, the transition from the angular stripe phase \rom{4} to the vortex-antivortex pair phase \rom{1} is continuous and the weighting coefficients $\alpha$ and $\gamma$ disappear gradually close to the transition.

In Fig.~\ref{fig1_phase}, we report the \emph{general} phase diagram at a given intra-spin interaction strength $g$ (a) or at a fixed ratio $g_{_{\uparrow\downarrow}}/g$ (b). At an experimentally accessible interaction strength $mg/\hbar^{2}\sim0.01$ in Fig.~\ref{fig1_phase}(a)~\cite{choi2013observation}, by decreasing $g_{_{\uparrow\downarrow}}/g$ from 1 we find that the angular stripe phase \rom{3} becomes favorable very soon, similar to the stripe phase in a Raman SOC Bose gas. In contrast, the angular stripe phase \rom{4} is less sensitive to the ratio and its parameter space is nearly unchanged upon deceasing $g_{_{\uparrow\downarrow}}/g$. At a typical ratio $g_{_{\uparrow\downarrow}}/g=0.5$ shown in Fig.~\ref{fig1_phase}(b), we examine the dependence of different phases on the overall interaction strength. With Raman SOC, it is known that a large interaction strength enhances the stripe and zero-momentum phases and suppresses the plane-wave phase, leading to a tri-critical point where three phases intervene~\cite{li2012quantum}. Here, we find that only the vortex-antivortex pair phase \rom{1} survives at large interaction strength. The large parameter space of the phase \rom{1} can be understood from the absence of the diagonal Stark shift in the recent experiments, which removes an additional confinement $\Omega(r)$ to both spin components~\cite{chen2018spin,zhang2019ground} and hence makes the zero-angular-momentum state $\psi_{_{0}}$ more energetically favorable by reducing the interaction energy.

{\color{blue}\emph{Collective modes and instabilities}}.\textemdash We now turn to discuss the low-lying collective excitations, which are readily measurable in SOC Bose gases~\cite{zhang2012collective,khamehchi2014measurement,ji2015softening}. In Figs.~\ref{fig3_collective}(a) and \ref{fig3_collective}(b), typical excitation spectra in the half-skyrmion phase \rom{2} and the vortex-antivortex pair phase \rom{1} are plotted as a function of the angular momentum $l$ of Bogoliubov quasiparticles, respectively. The Goldstone mode (i.e., condensate mode), low- and high-dipole modes (with $l=\pm\hbar$), and breathing mode (i.e, the lowest $l=0$ mode) are indicated by crosses, asterisks, circles and inverted triangles, respectively. In the phase \rom{2} we observe a clear roton structure, despite of the discreteness of the spectrum, similar to the roton found in the plane-wave phase of a Raman SOC Bose gas~\cite{martone2012anisotropic,khamehchi2014measurement,ji2015softening,zheng2012collective,chen2017quantum}. This originates from the spontaneous breaking of the axial symmetry and explains the first-order \rom{3}-\rom{2} phase transition we mentioned earlier.

In Fig.~\ref{fig3_collective}(c), a few low-lying mode frequencies are shown as a function of $\Omega$. The $\Omega$-dependence of the \emph{low-dipole} mode frequency (asterisks) is of particular interest and exhibits an intriguing behavior. By decreasing the Rabi frequency from the vortex-antivortex pair phase \rom{1}, we find that the low-dipole mode frequency becomes vanishingly small at $\Omega\simeq197.5\hbar\omega_{r}$, which is close to the \rom{1}-\rom{4} phase boundary $\Omega_{c}\simeq193.3\hbar\omega_{r}$ determined from the variational calculations. As $\Omega$ decreases further, the mode frequency shows a jump and then increases steadily, and finally saturates at about $0.62\omega_{r}$ at $\Omega=0$. This interesting $\Omega$-dependence of the dipole mode was observed earlier in a SOC Bose gas, where the complete softening of the mode occurs at the transition from the zero-momentum to plane-wave phases and is associated with the divergent effective mass or magnetic susceptibility at the transition~\cite{zhang2012collective,li2012sum,zheng2013properties}. In our case, although the mode frequency becomes discrete due to the quantization of angular momentum, the feature of complete softening remains. It is also a clear demonstration of the violation of Galilean invariance due to SOAMC~\cite{zheng2012collective}. On the other hand, the frequencies of the breathing mode (inverted triangles) and the high-dipole mode (circles) both experience a jump at the transition from the angular stripe phase \rom{4} to the phase \rom{2}, and sequentially increase upon decreasing $\Omega$. Near zero Rabi frequency, the breathing mode frequency is about $1.37\omega_{r}$, smaller than the scale-invariant classical prediction of $2\omega_{r}$~\cite{pitaevskii1997breathing} due to relatively large inter-atomic interactions. The high-dipole mode frequency approaches $\omega_{r}$, since the SOAMC can be gauged away in the $\Omega\to0$ limit and the Galilean invariance can be restored, ensuring the exact solution of Kohn mode with frequency $\omega_{r}$. At sufficiently large Rabi frequency, the mode frequencies in the same branch tend to approach each other since the low-lying excitation bands get flattened~\cite{demarco2015angular}.

We note that, close to the \rom{1}-\rom{4} transition a non-negligible imaginary part appears in the lowest excitation branch, as shown by the blue dotted curve in Fig.~\ref{fig3_collective}(d). This is simply the indication of the dynamical instability of the vortex-antivortex pair phase towards the phase transition. Furthermore, close to the \rom{3}-\rom{2} transition, the roton gap $\Delta$ (red dashed curve in Fig.~\ref{fig3_collective}(d)) starts to become negative, implying the energetic instability of the phase \rom{2} and thus determining a low bound for the \rom{3}-\rom{2} transition~\cite{chen2018quantum}. Two unstable regimes seen from the collective modes agree qualitatively with the red and blue shaded areas of the angular stripe phases \rom{4} and \rom{3} determined using the variational approach.

{\color{blue}\emph{Experimental detection}}.\textemdash To enlarge the parameter space of the novel angular stripe phases and to improve their visibility in current atomic experiments, we may utilize a Feshbach resonance (FR) at a
magnetic field $B_0=51.95$G of a SOAMC $^{41}$K gas in realistic harmonic traps with frequencies $(\omega_r,\omega_z)/2\pi=(10,200)$Hz. Near the FR, the intra-species scattering 
lengths $(a_{_{\uparrow\uparrow}},a_{_{\downarrow\downarrow}})\simeq(65,100)a_0$ are approximately constant, where $a_0$ is the Bohr radius, and the inter-species one $a_{_{\uparrow\downarrow}}$ can be tuned in a wide range~\cite{lysebo2010feshbach,tanzi2018feshbach}. In the upper panel of Fig.~\ref{fig4_n_correlation}, we show the phase diagram in the $\delta$-$\Omega$ plane by visualizing the spin magnetization $\langle\sigma_z\rangle$ at $a_{_{\uparrow\downarrow}}=10a_0$ (i.e., $B\approx51.83$G). 
It is readily seen that the stripe phase III occupies a relatively large window of $1.5\hbar\omega_r$ by tuning the detuning, which corresponds to a range of about $100$Hz in frequency and hence can be easily operated in experiments. To estimate the visibility in density profile, we introduce an angular density-density correlation function $g_{i}^{(2)}(\theta)\equiv\int_{0}^{2\pi}n_{i}(\phi)n_{i}(\phi+\theta)\mathrm{d}\phi/\int_{0}^{2\pi}n_{i}^{2}(\phi)\mathrm{d}\phi$ with the angular density $n_{i}(\phi)=\int_{0}^{\infty}r\mathrm{d}rn_{i}(r,\phi)$, and the label $i=\uparrow,\downarrow$ for each spin component and null for the total density. As shown in Fig.~\ref{fig4_n_correlation}(b) and Fig.~\ref{fig4_n_correlation}(d), the correlation $g^{(2)}(\theta)$ in the half-skyrmion phase II and the vortex-antivortex pair phase I is identically unity, due to the axial symmetry of the phases. In contrast, the novel angular stripe phases break the axial symmetry and exhibit spatial modulation in the directional angle $\theta$. In the stripe phase IV (Fig.~\ref{fig4_n_correlation}(c)), the modulation in $g^{(2)}(\theta)$ is relatively small due to the large portion of the $\psi_{_0}$ state in the condensate wavefunction, see also Fig.~\ref{fig2_coeff}(a). In sharp contrast, the angular stripe phase III shows a much larger spatial oscillation in $g^{(2)}(\theta)$ for both spin components, as can be seen in Fig.~\ref{fig4_n_correlation}(a). This hallmark feature might be useful in directly probing the existence of the angular stripe in experiments.

{\color{blue}\emph{Conclusions}}.\textemdash In summary, we have predicted the existence of two angular stripe phases in a Bose condensate with spin-orbital-angular-momentum coupling, which feature occupation of different angular-momentum states and have sizable spatial modulation in the angular density-density correlation. The phase space for these novel superfluids is notable and insensitive to the difference in the intra- and inter-species interactions. A Bose gas of $^{41}$K atoms in pancake traps could be a promising candidate system to probe the predicted stripe phase.

\begin{acknowledgments}
We acknowledge fruitful discussions with Yun Li and Ivana Vasi\'{c}. Our research was supported by the National Natural Science Foundation of China, Grant No. 11974384 (SGP) and No. 11804177 (PZ), the National Key Research and Development Program, Grant No. 2016YFA0301503 (SGP), the Shandong Provincial Natural Science Foundation, China, Grant No. ZR2018BA032 (PZ), and the Australian Research Council’s (ARC) Discovery Projects: FT140100003 and DP180102018 (XJL), FT130100815 and DP170104008 (HH).
\end{acknowledgments}

\bibliographystyle{apsrev4-1}
\bibliography{references}

\clearpage
\pagebreak
\widetext
\begin{center}
\textbf{\large{}Supplemental Materials: Angular Stripe Phase in Spin-Orbital-Angular-Momentum Coupled Bose Condensates}
\end{center}
\setcounter{equation}{0}
\setcounter{figure}{0}
\setcounter{table}{0}
\makeatletter
\renewcommand{\theequation}{S\arabic{equation}}
\renewcommand{\thefigure}{S\arabic{figure}}

\section{Single-particle dispersion and the variational approach}
We start with the single-particle Hamiltonian $\mathcal{H}_{s}$ given in Eq.~\eqref{eq:single-particle} of the main text in the $(r,\phi)$ plane. The axial symmetry makes the angular momentum $l_{z}$ a good quantum number, and we can calculate the single-particle states $\psi_{_{l_{z}}}^{(0)}$ as well as the dispersion relation $\varepsilon_{n}$ by employing the Schr\"{o}dinger equation \begin{equation} \label{eq:idealcase}
\mathcal{H}_{s}\psi_{_{l_{z}}}^{(0)}(r,\phi)=\varepsilon_{n}\psi_{_{l_{z}}}^{(0)}(r,\phi). 
\end{equation}

\begin{figure}[ht]
\centering
\includegraphics[width=0.96\textwidth]{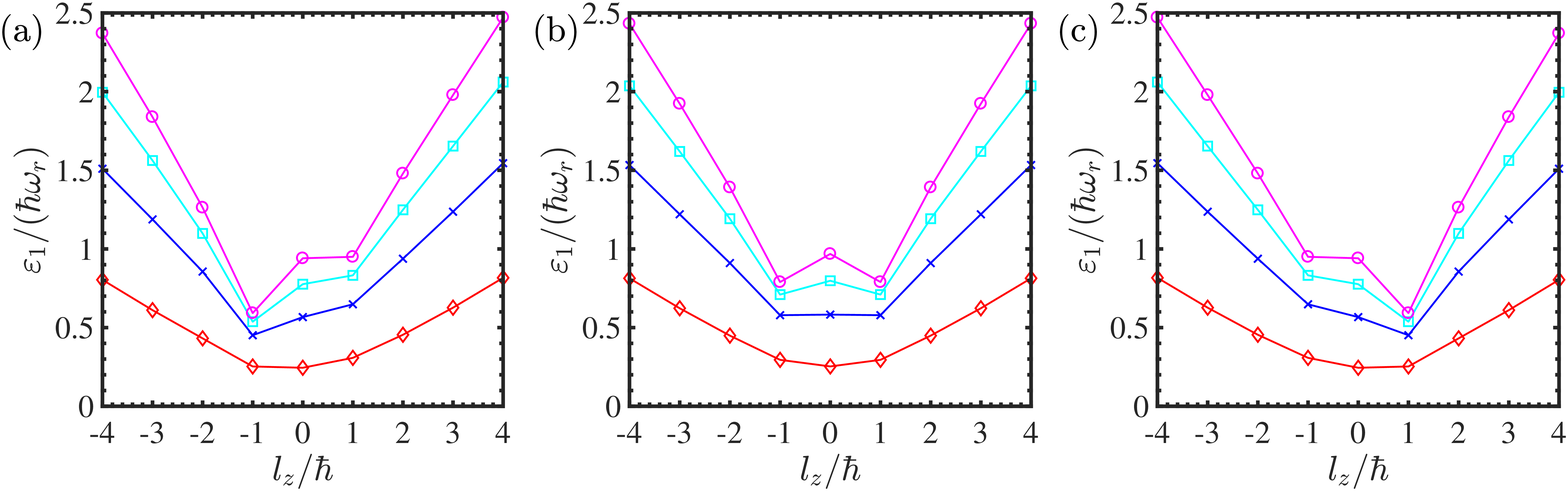}
\caption{The lowest branch $\varepsilon_{n=1}$ in the single-particle dispersion at the negative (a), zero (b) and positive (c) laser detuning $\delta$, as a function of the angular momentum $l_{z}$ when the Rabi frequency $\Omega$ increases (top to bottom).}
\label{figs1:spectrum_lz}
\end{figure}

In Fig.~\ref{figs1:spectrum_lz}, we depict the lowest energy level $\varepsilon_{n=1}(l_{z})$ as a function of $l_{z}$. At zero detuning $\delta=0$ (i.e., see the middle panel (b)), as the Rabi frequency $\Omega$ decreases (from bottom to top), the dispersion changes from the structure with a single-minimum at $l_{z}=0$ to that with two degenerate minima at $l_{z}=\pm\hbar$. Once the detuning of lasers are nonzero, the energy dispersion is asymmetric. As we decrease $\Omega$, the angular momentum of the condensate at the minimum changes from $l_{z}=0$ to $l_{z}=\hbar$ (or $-\hbar$) for $\delta>0$ (or $\delta<0$).

In the presence of interactions, the total Hamiltonian reads $\mathcal{H}=\mathcal{H}_{s}+\mathcal{H}_{\mathrm{int}}$. We may assume the angular momentum $l_{z}$ is still a good quantum number and determine the definite-angular-momentum (dam) states $\psi_{_{l_{z}}}^{(\mathrm{dam})}$ by self-consistently solving the nonlinear Schr\"{o}dinger equation or the Gross-Pitaevskii equation (see also Eq.~\eqref{eq:GP-eq-SOC} in the next section)
\begin{equation}  \label{eq:nonlinear}
    \mathcal{H}\psi^{(\mathrm{dam})}_{_{l_z}}(r,\phi)=\varepsilon_n\psi^{(\mathrm{dam})}_{_{l_z}}(r,\phi).
\end{equation}

In the variational approach, we can start with a variational \textit{ansatz} for the condensate wavefunction, i.e., a linear superposition of the lowest angular-momentum states $\psi_{_{l_{z}}}$, which can be either the single-particle states $\psi_{_{l_{z}}}^{(0)}$ or the self-consistently calculated definite-angular-momentum states $\psi_{_{l_{z}}}^{(\mathrm{dam})}$. The latter choice of course is more favorable, as the interaction effect is already included at the beginning. Explicitly, the condensate wavefunction is given by
\begin{equation} 
    {\bf \Psi}(r,\phi)=\alpha e^{i\theta_\alpha}\psi_{_{-1}}+\beta e^{i\theta_\beta}\psi_{_{0}}+\gamma e^{i\theta_\gamma}\psi_{_{1}},
\end{equation}
in terms of three phase angles $\theta_{\alpha}$, $\theta_{\gamma}$, $\theta_{\beta}$ and three real non-negative parameters $\alpha$, $\beta$, $\gamma$ satisfying the normalization condition $\alpha^{2}+\beta^{2}+\gamma^{2}=1$. The mean-field energy functional in terms of ${\bf \Psi}(r,\phi)\equiv\left(\Psi_{\uparrow},\Psi_{\downarrow}\right)^{T}$ is then written as
\begin{equation}
    E_\mathrm{mf}=\iint r\mathrm{d}r\mathrm{d}\phi \left[{\bf \Psi}^{\dagger}\mathcal{H}_{s}(\hat{L}_z){\bf \Psi}+\sum_{\sigma,\sigma^{\prime}}\frac{g_{\sigma\sigma^{\prime}}}{2}\Psi_{\sigma}^{\dagger}\Psi_{\sigma^{\prime}}^{\dagger}\Psi_{\sigma^{\prime}}\Psi_{\sigma}\right].
\end{equation}
In particular, the interaction energy per particle $\epsilon_{\textrm{int}}$
reads 
\begin{equation}
\epsilon_{\textrm{int}}=\frac{1}{2N}\iint r\mathrm{d}r\mathrm{d}\phi\left[g_{_{\uparrow\uparrow}}n_{\uparrow}n_{\uparrow}+g_{_{\downarrow\downarrow}}n_{\downarrow}n_{\downarrow}+2g_{_{\uparrow\downarrow}}n_{\uparrow}n_{\downarrow}\right],
\end{equation}
where we have introduced the density $n_{\sigma}(r,\phi)\equiv|\Psi_{\sigma}(r,\phi)|^{2}$ for spin component $\sigma=\uparrow,\downarrow$. The integral of the density terms $n_{\sigma}n_{\sigma^{\prime}}$ depends on the phase angles $\theta_{\alpha}$, $\theta_{\beta}$, and $\theta_{\gamma}$ through a cosine function $\cos{(\theta_{\alpha}-2\theta_{\beta}+\theta_{\gamma})}$. Therefore, the new angle $\Theta\equiv\theta_{\alpha}-2\theta_{\beta}+\theta_{\gamma}$ can take arbitrary value in the range $[0,2\pi]$, if the coefficient in front of the cosine function vanishes. Otherwise, $\Theta$ can only take values of $0$ or $\pi$ depending on the sign of the nonzero coefficient. Hence, the variational wavefunction of the condensed phase and relevant observables such as the spin polarizations can be obtained straightforwardly after the minimization of the energy functional with respect to these variational parameters.

\begin{figure}[ht]
\centering
\includegraphics[width=0.96\textwidth]{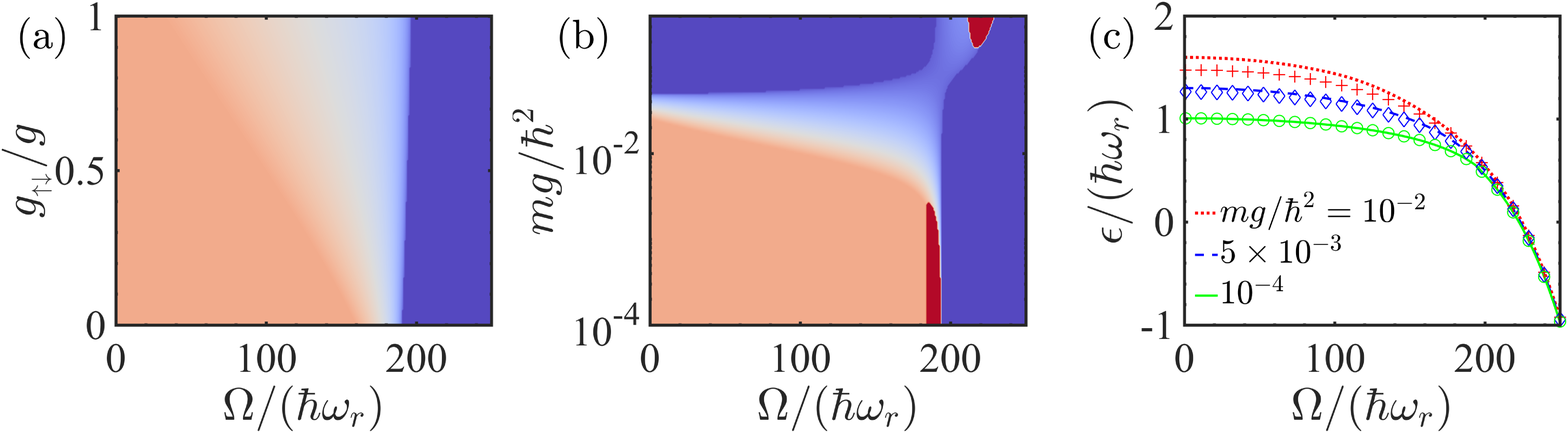}
\caption{(a) and (b): Phase diagram in the $g_{_{\uparrow\downarrow}}/g$-$\Omega$ plane and in the $g$-$\Omega$ plane, respectively, obtained by using the variational ansatz, which is constructed by using the single-particle states $\psi_{_{l_{z}}}^{(0)}$. The parameters are the same as in Fig.~\ref{fig1_phase}. The color indicates the portion of the $l_{z}=\hbar$ state in the condensate wavefunction. (c) The mean-field energy $\epsilon\equiv E_{\textrm{mf}}/N$ as a function of $\Omega$, at three sets of interaction strength $g$, calculated by using single-particle states (lines) and self-consistently-determined states (symbols) in the variational ansatz.}
\label{figs2:compareE}
\end{figure}

In Figs.~\ref{figs2:compareE}(a) and~\ref{figs2:compareE}(b), we show respectively the phase diagram in the $g_{_{\uparrow\downarrow}}/g$-$\Omega$ and $g$-$\Omega$ planes obtained by using the variational ansatz constructed from the single-particle states $\psi_{_{l_{z}}}^{(0)}$, which is to be compared with the ones constructed from the definite-angular-momentum states $\psi_{_{l_{z}}}^{(\mathrm{dam})}$ as shown in Fig.~1 in the main text. Two results have a quantitative agreement at sufficiently small interaction strength (i.e., $g\lesssim10^{-3}\hbar^{2}/m$). However, as the interactions become large, there are significant differences in the phase diagrams. These differences can be understood by comparing their mean-field energy $\epsilon$ as shown in Fig.~\ref{figs2:compareE}(c). The two energies agree very well at small $g$. Nonetheless, they start to become different as $g$ continues to increase. The difference is much more pronounced at small Rabi frequency: the energy calculated using $\psi_{_{l_{z}}}^{(\mathrm{dam})}$ (symbols) appears to be lower than the one using $\psi_{_{l_{z}}}^{(0)}$ (lines). Therefore, for the results discussed in the main text we adopt the variational ansatz constructed by using $\psi_{_{l_{z}}}^{(\mathrm{dam})}$.

\section{Some details on the Gross-Pitaevskii equation and the Bogoliubov theory}
To investigate the ground state at a \emph{definite} angular momentum $l_{z}$ and the corresponding collective behaviors, we generalize a Hartree-Fock-Bogoliubov theory with Popov approximation (HFBP)~\cite{griffin1996conserving,popov1991,chen2017quantum} and apply it to a two-dimensional SOAMC Bose gas at zero temperature. In a grand canonical ensemble, the Heisenberg equations of motion for the Bose operator field $\hat{\Phi}(r,\phi,t)$ take the form,
\begin{equation}  \label{eq:eom-soac}
\begin{aligned}
i\hbar\partial_t\hat{\Phi}_{\uparrow}&=[\mathcal{H}_{\mathrm{s}}^{(+)}-\mu]\hat{\Phi}_{\uparrow}+g\hat{\Phi}_{\uparrow}^{\dagger}\hat{\Phi}_{\uparrow}\hat{\Phi}_{\uparrow}+g_{_{\uparrow\downarrow}}\hat{\Phi}_{\downarrow}^{\dagger}\hat{\Phi}_{\downarrow}\hat{\Phi}_{\uparrow}+\Omega(r)\hat{\Phi}_{\downarrow},\\
i\hbar\partial_t\hat{\Phi}_{\downarrow}&=[\mathcal{H}_{\mathrm{s}}^{(-)}-\mu]\hat{\Phi}_{\downarrow}+g\hat{\Phi}_{\downarrow}^{\dagger}\hat{\Phi}_{\downarrow}\hat{\Phi}_{\downarrow}+g_{_{\uparrow\downarrow}}\hat{\Phi}_{\uparrow}^{\dagger}\hat{\Phi}_{\uparrow}\hat{\Phi}_{\downarrow}+\Omega(r)\hat{\Phi}_{\uparrow},
\end{aligned}
\end{equation}
with $\mathcal{H}_{\mathrm{s}}^{(\pm)}\equiv -\frac{\hbar^2}{2mr}\partial_r\left(r\partial_r\right)+\frac{(\hat{L}_z\mp n\hbar)^2}{2mr^2}+V_\mathrm{ext}(r)\pm\frac{\delta}{2}$ and the chemical potential $\mu$.

Following the standard procedure, the Bose field operator $\hat{\Phi}_{\sigma}(r,\phi,t)$ for spin component $\sigma=\uparrow,\downarrow$ can be rewritten as a combination of the condensate wavefunction $\Psi_{\sigma}$ and the non-condensate fluctuation operator $\hat{\eta}_{\sigma}$ as
\begin{equation}   \label{eq:newBosefield}
\begin{aligned}
    \hat{\Phi}_{\sigma}(r,\phi,t)=&\Psi_{\sigma}(r,\phi,t)+\hat{\eta}_{\sigma}(r,\phi,t)\\
    =&\Psi_{\sigma}(r,\phi)+\underset{j}{\sum}\left[u_{j\sigma}(r,\phi)e^{-i\varepsilon_jt}\hat{\alpha}_j+v^*_{j\sigma}(r,\phi)e^{i\varepsilon_jt}\hat{\alpha}_j^{\dagger}\right],
\end{aligned}
\end{equation} 
where we assume the static condensate wavefunction with a definite angular momentum $l_z$ as
\begin{equation} 
    \Psi_{\sigma}(r,\phi)= \varphi_\sigma(r)e^{il_z\phi}/\sqrt{2\pi},
\end{equation}
and the fluctuation operator $\hat{\eta}_{\sigma}$ is expanded in a quasiparticle basis ($\hat{\alpha}^{\dagger}$, $\hat{\alpha}$) under a Bogoliubov transformation. $\varepsilon_{j}$ is the quasiparticle frequency and the quasiparticle amplitudes $\mathcal{K}\equiv u$ (or $v$) in Eq.~\eqref{eq:newBosefield} can be written as $\mathcal{K}_{j\sigma}(r,\phi)=\mathcal{K}_{\tau\sigma}^{(l)}(r)r^{|l_{z}+l|}e^{i(l_{z}+l)\phi}/\sqrt{2\pi}$~\cite{vasic2016excitation}. Here the index $j$ is defined as $j\equiv(l,\tau)$ with the integer angular momentum $l\in\mathbb{Z}$ and the branch index $\tau\in\mathbb{N}_{1}$.

After substituting the new Bose field operator, Eq.~\eqref{eq:newBosefield}, into the equations of motion Eq.~\eqref{eq:eom-soac}, and applying the mean-field decoupling for the three-operator terms~\cite{griffin1996conserving}, we obtain two coupled equations.

(A) The first is the modified Gross-Pitaevskii (GP) equation for the static condensate
\begin{equation}  \label{eq:GP-eq-SOC}
\left[\mathcal{H}_{\mathrm{s}}(\hat{L}_z)+\mathrm{diag}(\mathcal{L}_\uparrow,\mathcal{L}_\downarrow)\right] {\bf \Psi}= \mu {\bf \Psi},               
\end{equation}
with the spinor ${\bf \Psi}\equiv\left(\Psi_{\uparrow},\Psi_{\downarrow}\right)^{T}$ and the diagonal element $\mathcal{L}_{\sigma}\equiv gn_{\sigma}+g_{_{\uparrow\downarrow}}n_{\bar{\sigma}}$ (here spin index $\sigma\neq\bar{\sigma}$).

(B) The other one is the coupled Bogoliubov equation for quasiparticles,
\begin{subequations}  \label{eq:Bogo-eq-SOAC}
\begin{eqnarray}
\left[\mathcal{H}_{s}(\hat{L}_z)-\mu+\mathcal{A}_\uparrow\right]U_j +\mathcal{B}V_j&=&\varepsilon_jU_j, \\
-\mathcal{B}U^*_j -\left[\mathcal{H}_{s}(\hat{L}_z)-\mu+\mathcal{A}^*_\downarrow\right]V^*_j&=&\varepsilon_jV^*_j,
\end{eqnarray}
\end{subequations}
where $U_j\equiv\left(u_{j\uparrow}, u_{j\downarrow}\right)^T$, $V_j\equiv\left(v_{j\uparrow}, v_{j\downarrow}\right)^T$, and
\begin{equation}\label{eq:matrix-AB}
\mathcal{A}_\sigma\equiv \left[
\begin{array}{cc}
2gn_{\uparrow}+g_{_{\uparrow\downarrow}}n_{\downarrow} &g_{_{\uparrow\downarrow}}\psi_{\sigma}\psi_{\bar{\sigma}}^*
\\ 
g_{_{\uparrow\downarrow}}\psi_{\bar{\sigma}}\psi_{\sigma}^* &2gn_{\downarrow}+g_{_{\uparrow\downarrow}}n_{\uparrow}
\end{array}
\right],~~~\mathcal{B}\equiv \left[
\begin{array}{cc}
g\psi^2_{\uparrow} &g_{_{\uparrow\downarrow}}\psi_{\uparrow}\psi_{\downarrow}
\\ 
g_{_{\uparrow\downarrow}}\psi_{\uparrow}\psi_{\downarrow} &g\psi^2_{\downarrow}
\end{array}
\right].
\end{equation}
Here $n_{\sigma}=|\psi_{\sigma}|^{2}$ denotes the density for spin component $\sigma$ with $\sigma\neq\bar{\sigma}$. Note that, in the GP and Bogoliubov equations, the noncondensate density is neglected at zero temperature since the quantum depletion is typically negligible.

Therefore, the wavefunction or relevant energy of the states with different angular momentum can be obtained by self-consistently solving the GP equation. After numerically calculating the ground-state wavefunction ${\bf \Psi}(r,\phi)$, one obtains straightforwardly the elementary excitations $\varepsilon_{j}$ with respect to the quasiparticle angular momentum $l$ via the Bogoliubov equations in Eq.~\eqref{eq:Bogo-eq-SOAC}. The typical spectrum in the phases \rom{1} and \rom{2} can be seen in Figs.~\ref{fig3_collective}(a) and \ref{fig3_collective}(b), while the behaviors of collective-mode frequencies as a function of Rabi frequency are depicted in Figs.~\ref{fig3_collective}(c) and \ref{fig3_collective}(d).

\end{document}